# SUSY Invariants from the BRST Cohomology of the SOSO Model

JOHN A. DIXON[*]
MATHEMATICAL INSTITUTE,
OXFORD UNIVERSITY
OXFORD, ENGLAND

**Abstract**

The SOSO action is an irreducible action for a complex massive superspin one half representation of SUSY, made from reality constrained spin one half superfields. The theory requires 'BRST recycling' to find appropriate nilpotent BRST transformations. A superfield treatment is probably not available, which means that mass terms, possible anomalies, and couplings to other representations, are all to be found using BRST cohomology. In two earlier papers two mass terms and a potential anomaly were examined, without explaining how they arose from the BRST cohomology. This paper is designed to fill that gap in the theory.

**1**. For a long time it has been evident that the BRST cohomology of the chiral scalar superfield seems to naturally couple to a chiral dotted spinor superfield [19,20,21,22]. But a reasonable action for the chiral dotted spinor superfield was hard to find. The most obvious action had higher derivatives and tachyons [17]. It also contained a vector boson. Progress on SUSY representations of massive higher superspin actions and their field equations [30,31] helped to focus on ideas to resolve these problems. The result was that a sensible action for this kind of superfield was found, but it required 'BRST Recycling' [17] and it needed to be formulated in terms of BRST cohomology. This meant that the action was written in terms of component fields, probably without any possible superfield version. We will call the model of [17,18] the SOSO model, since it needs a short name[2].

**2**. In [17] it was shown that there was a mass term, and that the theory describes a complex irreducible massive superspin one half theory. In [18] a second mass term was added. It was called an Extraordinary Mass Invariant because it contains Zinn sources. Then it was shown

---

[*]jadixg@gmail.com;
dixon@maths.ox.ac.uk

[2]SOSO is a pronounceable acronym which picks out four letters from the more accurate but very long description: ' Complex Massive **S**uperspin **O**ne Half Action with an overall conserved global phase, made from two 'reality constrained' pairs of Chiral Dotted and Undotted **S**pin **O**ne Half Superfields, using BRST recycling to derive the BRST transformations of the Dotted Chiral Spinor Superfields from those of the Undotted Chiral Spinor Superfields'. This was explained in detail in [17,18]

that adding both types of mass terms still makes sense so long as the action is completed with a vector mass term, and also constrained so that the coefficient of the completion obstruction is zero [18].

**3.** This paper will explain how the mass terms and the obstruction terms can be found using BRST cohomology. The computation of BRST cohomology has a long history. For general surveys, see, for example, [1,2,3,4,5,6,7,8,9]. For BRST cohomology of supersymmetric theories, in particular, see, for example, [10,11,12,13,14,19,20,21,22]. This paper follows on from that tradition, but some effort will be made to make this paper fairly self contained.

**4.** It is well known that SUSY is the 'square root' of a relativistic translation in space and time [32]. In terms of BRST this means that it is governed by the nilpotent super Lie algebra boundary operator:

$$\delta_\xi = -C_\alpha \overline{C}_{\dot\alpha} \frac{\partial}{\partial \xi_{\alpha\dot\alpha}} \equiv -C_\alpha \overline{C}_{\dot\alpha} \xi^\dagger_{\alpha\dot\alpha} \tag{1}$$

where $C, \overline{C}$ are 'commuting' spinors and $\xi$ is an 'anticommuting' translation.

**5.** The actual calculation of the cohomology here is quite simple. The first step is to construct all possible terms made from some basic building blocks, with some rules. This space is called $E_1$, and it divides into a number of simple subspaces. Here is an example of a term in $E_1$

$$m^2 (C^\alpha \xi_{\alpha\dot\alpha} \overline{C}^{\dot\alpha}) \omega \overline{\omega} \tag{2}$$

The building blocks and rules will be discussed below. For the case of this particular example (2), this is the end of the story, and it represents an object[3] in the Cohomology space $\mathcal{H}$.

**6.** The operator to be analyzed here is defined by the action $\mathcal{A}_{\text{Massless}}$ discussed[4] in [18]. It has the form:

$$\mathcal{A}_{\text{Massless}} = \mathcal{A}_{\text{Kinetic }\chi} + \mathcal{A}_{\text{Kinetic }\phi} + \mathcal{A}_{\text{Zinn }\chi} + \mathcal{A}_{\text{Zinn }\phi} + \mathcal{A}_{\text{GGF}} + \mathcal{A}_{\text{SUSY}} \tag{3}$$

where the component actions were introduced in [17]. Using the BRST Poisson Bracket defined by (28) below[5], one can write down the operator $\delta_{\text{Massless}} \equiv \delta$ that we will analyze in this paper. This operator is written in detail below in equations (35) to (59). The technique outlined in [15] and the results set out in [16] form the basis of the calculation here. The main new development here, besides the SOSO model, is the choice of Grading, which is set down in equation (18).

**7.** The expressions $\mathcal{A}_O, \mathcal{A}_E$ and $\mathcal{O}$ that were used in [17,18] are indeed in the cohomology space $\mathcal{H}$ of $\delta_{\text{Massless}}$. They can be obtained[6] from the following maps[7]:

$$m(C\xi\overline{C})E\overline{\omega} \in E_\infty \Rightarrow \mathcal{A}_O \in \mathcal{H} \tag{7}$$

---

[3] See equation (9).

[4] In fact that action did not include the ghost and gauge fixing action $\mathcal{A}_{\text{GGF}}$, but that makes no difference to the result for the cohomology, as can be seen below.

[5] This comes from equation (6) of [17]

[6] See (79) (86), (87), (89), (90) and (92).

[7] For example the object $\mathcal{A}_E$ in equation (8) above has the full form [18]:

$$\mathcal{A}_E = \int d^4x \left\{ 2m_2 \Upsilon \overline{\omega} - \frac{m_2}{2} \partial_{\alpha\dot\alpha} \overline{V}^{\alpha\dot\alpha} E - m_2 Z_L^{\dot\alpha} C^\alpha \overline{V}_{\alpha\dot\alpha} \right. \tag{4}$$

$$m(C\xi\overline{C})J'\overline{\omega} \in E_\infty \Rightarrow \mathcal{A}_{\mathrm{E}} \in \mathcal{H} \tag{8}$$

$$m^2(C\xi\overline{C})\omega\overline{\omega} \in E_\infty \Rightarrow \mathcal{O} \in \mathcal{H} \tag{9}$$

The space $E_\infty$ is isomorphic [15] to the cohomology space $\mathcal{H}$, as shown in Equation (26), and the map from $E_\infty$ to $\mathcal{H}$ is explained after equation (118). Some additional results, not used in [17,18], are that there is another Extraordinary Mass Invariant $\mathcal{A}_{\mathrm{E}2}$, and there are also two more Obstructions $\mathcal{O}_2, \mathcal{O}_3$ of ghost charge one (also their complex conjugates):

$$m(C\xi^2 C)(E+2J')(\overline{E}-2\overline{J'}) \in E_\infty \Rightarrow \mathcal{A}_{\mathrm{E}2} \in \mathcal{H} \tag{10}$$

$$m^2(C\xi^2 C)(E+2J')\overline{\omega} \in E_\infty \Rightarrow \mathcal{O}_2 \in \mathcal{H} \tag{11}$$

$$m^2(C\xi^2 C)(\overline{E}-2\overline{J'})\omega, \in E_\infty \Rightarrow \mathcal{O}_3 \in \mathcal{H} \tag{12}$$

This paper will not discuss the explicit forms of $\mathcal{A}_{\mathrm{E}2}$, $\mathcal{O}_2$ and $\mathcal{O}_3$ that arise from (10), (11) and (12) here. These do need to be understood, of course, along the lines of [18].

**8**. Here are tables of all the pairs of fields and Zinns and other symbols in this model, with their mass Dimension Dim, Lepton number $N_{\mathrm{L}}$ and their 'Ghost Number' $N_{\mathrm{G}}$.

| $\chi$ Fields, Ghosts, Zinns | | | | $\phi$ Fields, Ghosts, Zinns | | | |
|---|---|---|---|---|---|---|---|
| Field | Dim | $N_{\mathrm{L}}$ | $N_{\mathrm{G}}$ | Field | Dim | $N_{\mathrm{L}}$ | $N_{\mathrm{G}}$ |
| $\chi_{L\dot\alpha}$ | $\frac{3}{2}$ | 1 | 0 | $\phi_{L\dot\alpha}$ | $\frac{3}{2}$ | 1 | 0 |
| $U_{R\dot\alpha}$ | $\frac{5}{2}$ | -1 | -1 | $Z_{R\dot\alpha}$ | $\frac{5}{2}$ | -1 | -1 |
| $\chi_{R\dot\alpha}$ | $\frac{3}{2}$ | -1 | 0 | $\phi_{R\dot\alpha}$ | $\frac{3}{2}$ | -1 | 0 |
| $U_{L\dot\alpha}$ | $\frac{5}{2}$ | 1 | -1 | $Z_{L\dot\alpha}$ | $\frac{5}{2}$ | 1 | -1 |
| $V_{\alpha\dot\alpha}$ | 1 | 1 | 0 | $W_{\alpha\dot\alpha}$ | 2 | 1 | 0 |
| $\overline{\Omega}_{\alpha\dot\alpha}$ | 3 | -1 | -1 | $\overline{\Sigma}_{\alpha\dot\alpha}$ | 2 | -1 | -1 |
| $B$ | 2 | 1 | 0 | $E$ | 1 | 1 | 0 |
| $\overline{\Xi}$ | 2 | -1 | -1 | $\overline{\Upsilon}$ | 3 | -1 | -1 |
| $\omega$ | 0 | 1 | 1 | $\omega'$ | 0 | 1 | 1 |
| $\overline{K}$ | 4 | -1 | -2 | $\overline{K}'$ | 4 | -1 | -2 |
| $\eta$ | 2 | 1 | -1 | $\eta'$ | 3 | 1 | -1 |
| $\overline{J}$ | 2 | -1 | 0 | $\overline{J}'$ | 1 | -1 | 0 |
| $L$ | 2 | 1 | 0 | $L'$ | 2 | 1 | 0 |
| $\overline{\Delta}$ | 2 | -1 | -1 | $\overline{\Delta}'$ | 2 | -1 | -1 |

(13)

The complex conjugate quantities of the objects in the tables above have the opposite signs for 'Lepton number' $N_{\mathrm{L}}$, and the same signs for 'Ghost number' $N_{\mathrm{G}}$ and dimension Dim.

---

$$+ m_2 \overline{Z}^\alpha_R \overline{C}^{\dot\alpha} V_{\alpha\dot\alpha} + m_2 \phi_{L\dot\alpha}\chi_R^{\dot\alpha} - m_2 \overline{\phi}_{R\alpha}\overline{\chi}_L^\alpha - m_2 \Sigma^{\alpha\dot\alpha}\overline{C}_{\dot\beta}\overline{\chi}_{L\alpha} + m_2 \Sigma^{\alpha\dot\alpha}\chi_{R\dot\alpha}C_\alpha + 2m_2 J'\overline{B}\} + * \tag{5}$$

Of course this expression is not unique—one can add to it any local boundary. So one can change the form of the above to

$$\mathcal{A}_{\mathrm{E}} \Rightarrow \mathcal{A}_{\mathrm{E}} + \delta\mathcal{B} + *; \quad \mathcal{B} = \int d^4 x \left\{ b_1 \overline{\Sigma}^{\alpha\dot\alpha} V_{\alpha\dot\alpha} + b_2 \overline{\eta} E \right\} \tag{6}$$

Here are the values of these for some other expressions:

$$\text{Dim } C_\alpha = \text{Dim } \overline{C}_{\dot\alpha} = -\frac{1}{2}; \text{ Dim } \xi_{\alpha\dot\alpha} = -1 \qquad (14)$$

$$N_G\, C_\alpha = N_G\, \overline{C}_{\dot\alpha} = N_G\, \xi_{\alpha\dot\alpha} = 1 \qquad (15)$$

$$\text{Dim } m = \text{Dim } \partial = 1; N_G\, m = N_G\, \partial = 0 \qquad (16)$$

Usually in gauge theory, the ghost number of the action is taken to be zero, but because of the way the spectral sequence works, it is more convenient to take it to be 4:

$$\text{Dim} \int d^4 x = -4; \ N_G \int d^4 x = 4 \qquad (17)$$

These assignmnents ensure that the action $\mathcal{A}$ has dimension $\text{Dim } \mathcal{A} = 0$, Ghost number $N_G\, \mathcal{A} = 4$ and Lepton number $N_L\, \mathcal{A} = 0$. This causes no trouble, and we will 'renormalize' the ghost number to its conventional value of zero for the action, and one for an anomaly or obstruction, when it is convenient.

**9**. A spectral sequence is entirely determined once one has chosen the operator $\delta$ and the Grading. Here is the Grading that will be used here:

$$N_{\text{Grading}} = N_C + N_{\overline{C}} + 2 N_\xi \qquad (18)$$

Given the above grading the operator decomposes into a sum:

$$\delta = \delta_0 + \delta_1 + \delta_2, \ [N_{\text{Grading}}, \delta_i] = i \delta_i \qquad (19)$$

Then the operator $\delta_0$ consist of three parts:

$$\delta_0 = \delta_{0\chi} + \delta_{0\phi} + \delta_\xi \qquad (20)$$

where $\delta_{0\chi}$ is in (65) through (67), and $\delta_{0\phi}$ is in (70) through (71), and $\delta_\xi$ is in (1). The operator $\delta_2$ has the form:

$$\delta_2 = \xi^{\alpha\dot\alpha} \partial_{\alpha\dot\alpha} + \int d^4 x \left( V^{\beta\dot\beta} C_\beta \overline{C}_{\dot\beta} \frac{\delta}{\delta\omega} + K \overline{C}_{\dot\alpha} C_\alpha \frac{\delta}{\delta\Omega_{\alpha\dot\alpha}} \right. \qquad (21)$$

$$\left. -\frac{1}{2} C_\beta \overline{C}_{\dot\beta} \partial^{\beta\dot\beta} \eta \frac{\delta}{\delta J} + \eta' \overline{C}_{\dot\alpha} C_\alpha \frac{\delta}{\delta W_{\alpha\dot\alpha}} - \Sigma^{\beta\dot\beta} C_\beta \overline{C}_{\dot\beta} \frac{\delta}{\delta J'} \right) + * \qquad (22)$$

where these terms come from equations (59) and then (39), (44), (47), (51) and (58). The remaining operator $\delta_1$ is what remains from (35) to (59) after the above have been removed. It has the form $\delta_1 = C^\alpha \nabla_\alpha + \overline{C}^{\dot\alpha} \overline{\nabla}_{\dot\alpha}$. We shall pick out the important parts of it, for the present calculation, in equation (74).

**10**. The counting operator $N_\phi$ counts the number of $\phi$ fields and Zinns (as listed in Table 13). It commutes with $\delta$. Thus, using obvious definitions:

$$[N_\phi, \delta] = [N_\chi, \delta] = [N_L, \delta] = [N_{\text{Dim}}, \delta] = [N_m, \delta] \qquad (23)$$

$$= [N_\phi, d_r] = [N_\chi, d_r] = [N_L, d_r] = [N_{\text{Dim}}, d_r] = [N_m, d_r] = 0 \tag{24}$$

for all the $d_r$, where the $d_r$ are as defined in [15]. It is shown in [15] that:

$$[N_G, \delta] = \delta; \ [N_G, d_r] = d_r; \ [N_{\text{Grading}}, d_r] = r d_r \tag{25}$$

$$E_{r+1} = \ker\left(d_r \cap d_r^\dagger\right) \cap E_r; \ E_\infty \subset \cdots E_{r+1} \subset E_r \cdots \subset E_0; \ \mathcal{H} \approx E_\infty \tag{26}$$

The objects in $E_r$ have Dim $E_r = 0$. Also we have Dim $\mathcal{H} = 0$. $E_0$ is the space we start with. It consists of unintegrated polynomials in the Fields, Zinns, Ghosts and derivatives and the parameters $\xi, C, \overline{C}$. The spectral sequence is said to collapse when $d_r$ becomes zero:

$$d_r = 0 \text{ for } r \geq n \Leftrightarrow E_n = E_{n+1} = \cdots = E_\infty \approx \mathcal{H} \tag{27}$$

In the sectors we examine in the present paper we find $d_2 = d_3 \cdots = 0$ and so $E_2 = E_\infty$. The Ghost number and other conserved quantum numbers of objects in $E_r$ and of their images under the mapping to $E_\infty \to \mathcal{H}$ are matched. Of course the parts with $N_G = 0, 1$ are of particular interest[8] [29].

**11**. Here is the BRST Poisson Bracket:

$$\mathcal{P}_{\text{Total}}[\mathcal{A}] = \mathcal{P}_\chi[\mathcal{A}] + \mathcal{P}_\phi[\mathcal{A}] + \mathcal{P}_{\text{SUSY}}[\mathcal{A}] \tag{28}$$

where

$$\mathcal{P}_\chi[\mathcal{A}] = \int d^4 x \left\{ \frac{\delta \mathcal{A}}{\delta U_{R\dot\alpha}} \frac{\delta \mathcal{A}}{\delta \chi_L^{\dot\alpha}} + \frac{\delta \mathcal{A}}{\delta U_{L\dot\alpha}} \frac{\delta \mathcal{A}}{\delta \chi_R^{\dot\alpha}} \right. \tag{29}$$

$$\left. + \frac{\delta \mathcal{A}}{\delta \overline{\Omega}_{\alpha\dot\alpha}} \frac{\delta \mathcal{A}}{\delta V^{\alpha\dot\alpha}} + \frac{\delta \mathcal{A}}{\delta \overline{\Xi}} \frac{\delta \mathcal{A}}{\delta B} + \frac{\delta \mathcal{A}}{\delta \overline{K}} \frac{\delta \mathcal{A}}{\delta \omega} + \frac{\delta \mathcal{A}}{\delta \overline{J}} \frac{\delta \mathcal{A}}{\delta \eta} + * \right\} \tag{30}$$

$$\mathcal{P}_\phi[\mathcal{A}] = \int d^4 x \left\{ \frac{\delta \mathcal{A}}{\delta Z_L^{\dot\alpha}} \frac{\delta \mathcal{A}}{\delta \phi_{R\dot\alpha}} + \frac{\delta \mathcal{A}}{\delta Z_R^{\dot\alpha}} \frac{\delta \mathcal{A}}{\delta \phi_{L\dot\alpha}} \right. \tag{31}$$

$$\left. + \frac{\delta \mathcal{A}}{\delta \overline{\Sigma}^{\alpha\dot\alpha}} \frac{\delta \mathcal{A}}{\delta W_{\alpha\dot\alpha}} + \frac{\delta \mathcal{A}}{\delta \overline{\Upsilon}} \frac{\delta \mathcal{A}}{\delta E} + \frac{\delta \mathcal{A}}{\delta \overline{J}'} \frac{\delta \mathcal{A}}{\delta \eta'} + * \right\} \tag{32}$$

$$\mathcal{P}_{\text{SUSY}}[\mathcal{A}] = \frac{\partial \mathcal{A}}{\partial h_{\alpha\dot\alpha}} \frac{\partial \mathcal{A}}{\partial \xi^{\alpha\dot\alpha}} \tag{33}$$

It should be noted that the above operator has been modified from the one in [17] by dropping some irrelevant terms. The terms $\frac{\delta \mathcal{A}}{\delta \overline{\Delta}} \frac{\delta \mathcal{A}}{\delta L} + *$ have been dropped because there is an integration over the auxiliary $L$ here. The terms $\frac{\delta \mathcal{A}}{\delta \omega'} \frac{\delta \mathcal{A}}{\delta K'} + \frac{\delta \mathcal{A}}{\delta \overline{\Delta}'} \frac{\delta \mathcal{A}}{\delta L'} + *$ have been dropped because they play no role, and are decoupled. The gauge fixing and ghost action from [17], including the Zinn term with $J$ and the gauge fixing part from $\delta_{\text{Massless}} \overline{\eta}$, after completing the quadratic forms and integrating over $L$ yields

$$\mathcal{A}_{\text{GF plus J Zinn}} = -\int d^4 x \frac{1}{2g} \left( \left[\partial_{\alpha\dot\alpha} \overline{V}^{\alpha\dot\alpha} + 2\overline{J}\right] \left[\partial_{\beta\dot\beta} V^{\beta\dot\beta} + 2J\right] \right) \tag{34}$$

---

[8]In the notation here these are actually $N_G = 4, 5$ unless we redefine them, which we generally do for integrated expressions, to be consistent with the usual treatment in gauge theory.

and we use that below for part of the action to take the relevant functional derivatives. All the other necessary terms are in [17].

**12**. Here is a complete version of all the BRST transformations in the operator $\delta \equiv \delta_{\text{Massless}}$. It arises from the action $\mathcal{A}_{\text{Massless}}$ defined in equation (3) by taking the 'square root' of the BRST Poisson Bracket above.

The $\chi$ Field, Ghost and Antighost transformations come from the Zinn action in equations (32) to (38) of [17] and equation (34) above:

$$\delta \chi_L^{\dot\alpha} = \frac{\delta \mathcal{A}_{\text{Massless}}}{\delta U_{R\dot\alpha}} = \left( B\overline{C}^{\dot\alpha} + G^{(\dot\alpha\dot\beta)}\overline{C}_{\dot\beta} \right) \tag{35}$$

$$\delta \chi_R^{\dot\alpha} = \frac{\delta \mathcal{A}_{\text{Massless}}}{\delta U_{L\dot\alpha}} = \left( -\overline{B}\overline{C}^{\dot\alpha} + \overline{G}^{(\dot\alpha\dot\beta)}\overline{C}_{\dot\beta} \right) \tag{36}$$

$$\delta \overline{B} = \frac{\delta \mathcal{A}_{\text{Massless}}}{\delta \Xi} = \left( \frac{1}{2}\partial^{\alpha\dot\alpha}\overline\chi_{L\alpha}\overline{C}_{\dot\alpha} - \frac{1}{2}\partial^{\alpha\dot\alpha}\chi_{R\dot\alpha}C_\alpha \right) \tag{37}$$

$$\delta \overline{V}^{\alpha\dot\alpha} = \frac{\delta \mathcal{A}_{\text{Massless}}}{\delta \Omega_{\alpha\dot\alpha}} = \left( \partial^{\alpha\dot\alpha}\overline\omega + \chi_R^{\dot\alpha}C^\alpha + \overline\chi_L^\alpha \overline{C}^{\dot\alpha} \right) \tag{38}$$

$$\delta \overline\omega = \frac{\delta \mathcal{A}_{\text{Massless}}}{\delta K} = \left( \overline{V}^{\beta\dot\beta}C_\beta \overline{C}_{\dot\beta} \right) \tag{39}$$

$$\delta \eta = \frac{\delta \mathcal{A}_{\text{Massless}}}{\delta \overline{J}} = -\frac{1}{g}\left[ \partial_{\alpha\dot\alpha}V^{\alpha\dot\alpha} + 2J \right] \tag{40}$$

The $\chi$ Zinns transformations come from the kinetic action $\mathcal{A}_{\text{Kinetic }\chi}$ in equation (24) in [17], and the ghost action[9] $\mathcal{A}_{\text{G}}$ in equation (72) in [17], and from the Zinn action $\mathcal{A}_{\text{Zinn }\chi \text{ Form 2}}$ in equations (39) to (46) in [17] and from equation (34):

$$\delta U_{R\dot\alpha} = \frac{\delta \mathcal{A}_{\text{Massless }\chi}}{\delta \chi_L^{\dot\alpha}} = \partial_{\alpha\dot\alpha}\overline\chi_L^\alpha + \frac{1}{2}\partial_{\alpha\dot\alpha}\Xi C^\alpha - \overline\Omega_{\gamma\dot\alpha}C^\gamma \tag{41}$$

$$\delta U_{L\dot\alpha} = \frac{\delta \mathcal{A}_{\text{Massless }\chi}}{\delta \chi_R^{\dot\alpha}} = \partial_{\alpha\dot\alpha}\overline\chi_R^\alpha - \frac{1}{2}\partial_{\alpha\dot\alpha}\Xi C^\alpha - \Omega_{\alpha\dot\alpha}C^\alpha \tag{42}$$

$$\delta \overline\Omega_{\alpha\dot\alpha} = \frac{\delta \mathcal{A}_{\text{Massless }\chi}}{\delta V^{\alpha\dot\alpha}} = \Box \overline{V}_{\alpha\dot\alpha} + \frac{1}{2}\partial_{\alpha\dot\alpha}\partial^{\beta\dot\beta}\overline{V}_{\beta\dot\beta} + \partial_{\alpha\dot\alpha}\frac{1}{2g}\left[ \partial_{\beta\dot\beta}\overline{V}^{\beta\dot\beta} + 2\overline{J} \right] \tag{43}$$

$$+ \overline{K}C_{\dot\alpha}C_\alpha - \frac{1}{2}\partial_\alpha^{\dot\gamma}U_{R\dot\gamma}\overline{C}_{\dot\alpha} - \frac{1}{2}\partial_\alpha^{\dot\gamma}U_{R\dot\alpha}\overline{C}_{\dot\gamma} - \frac{1}{2}\partial_{\dot\alpha}^{\gamma}\overline{U}_{L\gamma}C_\alpha - \frac{1}{2}\partial_{\dot\alpha}^{\gamma}\overline{U}_{L\alpha}C_\gamma \tag{44}$$

$$\delta \Xi = \frac{\delta \mathcal{A}_{\text{Massless}}}{\delta \overline{B}} = -2B + \overline{U}_{R\beta}C^\beta - U_{L\dot\beta}\overline{C}^{\dot\beta} \tag{45}$$

---

[9]Here we have corrected a missing factor of $\frac{1}{2}$ in the $\eta\overline\eta$ term of equations (72) and (94) of [17].

$$\delta \overline{K} = \frac{\delta \mathcal{A}_{\text{Massless }\chi}}{\delta \omega} = -\Box \overline{\eta} + \overline{\Omega}_{\gamma \dot{\delta}} \tag{46}$$

$$\delta \overline{J} = \frac{\delta \mathcal{A}_{\text{Massless}}}{\delta \eta} = \Box \overline{\omega} - \frac{1}{2} g C_\beta \overline{C}_{\dot{\beta}} \partial^{\beta \dot{\beta}} \overline{\eta} - \frac{1}{2} \partial_{\alpha \dot{\alpha}} \left( \chi_R^{\dot{\alpha}} C^\alpha + \overline{\chi}_L^\alpha \overline{C}^{\dot{\alpha}} \right) \tag{47}$$

The $\phi$ Field and Antighost transformations come from the Zinn action $\mathcal{A}_{\text{Zinn }\phi}$ in equations (47) to (54) of [17].

$$\delta \phi_{L \dot{\alpha}} = \frac{\delta \mathcal{A}_{\text{Massless}}}{\delta Z_R^{\dot{\alpha}}} = -\frac{1}{2} \partial_{\alpha \dot{\alpha}} E C^\alpha - W_{\alpha \dot{\alpha}} C^\alpha \tag{48}$$

$$\delta \phi_{R \dot{\alpha}} = \frac{\delta \mathcal{A}_{\text{Massless}}}{\delta Z_L^{\dot{\alpha}}} = \frac{1}{2} \partial_{\alpha \dot{\alpha}} \overline{E} C^\alpha - \overline{W}_{\alpha \dot{\alpha}} C^\alpha \tag{49}$$

$$\delta E = \frac{\delta \mathcal{A}_{\text{Massless}}}{\delta \overline{\Upsilon}} = \overline{\phi}_{R \beta} C^\beta - \phi_{L \dot{\beta}} \overline{C}^{\dot{\beta}} \tag{50}$$

$$\delta W_{\alpha \dot{\alpha}} = \frac{\delta \mathcal{A}_{\text{Massless}}}{\delta \overline{\Sigma}^{\alpha \dot{\alpha}}} = \eta' C_\alpha \overline{C}_{\dot{\alpha}} - \frac{1}{2} \partial_\alpha^{\dot{\gamma}} \phi_{L \dot{\gamma}} \overline{C}_{\dot{\alpha}} - \frac{1}{2} \partial_\alpha^{\dot{\gamma}} \phi_{L \dot{\alpha}} \overline{C}_{\dot{\gamma}} \tag{51}$$

$$- \frac{1}{2} \partial_{\dot{\alpha}}^\gamma \overline{\phi}_{R \gamma} C_\alpha - \frac{1}{2} \partial_{\dot{\alpha}}^\gamma \overline{\phi}_{R \alpha} C_\gamma \tag{52}$$

$$\delta \overline{\eta}' = \frac{\delta \mathcal{A}_{\text{Massless}}}{\delta J'} = \partial^{\alpha \dot{\alpha}} \overline{W}_{\alpha \dot{\alpha}} \tag{53}$$

The $\phi$ Zinn transformations come from the action $\mathcal{A}_{\text{Kinetic }\phi}$ in equation (29) of [17], and from the Zinn action $\mathcal{A}_{\text{Zinn }\phi\text{ Form 2}}$ in equations (58) to (63) of [17].

$$\delta Z_L^{\dot{\alpha}} = \frac{\delta \mathcal{A}_{\text{Massless}}}{\delta \phi_{R \dot{\alpha}}} = \partial^{\alpha \dot{\alpha}} \overline{\phi}_{R \alpha} + \frac{1}{2} \eta' \overline{C}^{\dot{\alpha}} - \Upsilon \overline{C}^{\dot{\alpha}} - \widetilde{\Sigma}^{(\dot{\alpha} \dot{\beta})} \overline{C}_{\dot{\beta}} \tag{54}$$

$$\delta Z_R^{\dot{\alpha}} = \frac{\delta \mathcal{A}_{\text{Massless}}}{\delta \phi_{L \dot{\alpha}}} = \partial^{\alpha \dot{\alpha}} \overline{\phi}_{L \alpha} + \frac{1}{2} \overline{\eta}' \overline{C}^{\dot{\alpha}} + \overline{\Upsilon} \overline{C}^{\dot{\alpha}} - \widetilde{\overline{\Sigma}}^{(\dot{\alpha} \dot{\beta})} \overline{C}_{\dot{\beta}} \tag{55}$$

$$\delta \overline{\Upsilon} = \frac{\delta \mathcal{A}_{\text{Massless}}}{\delta E} = -\frac{1}{2} \Box \overline{E} - \frac{1}{2} \partial^{\alpha \dot{\alpha}} \overline{Z}_{L \alpha} \overline{C}_{\dot{\alpha}} + \frac{1}{2} \partial^{\alpha \dot{\alpha}} Z_{R \dot{\alpha}} C_\alpha \tag{56}$$

$$\delta \overline{\Sigma}^{\alpha \dot{\alpha}} = \frac{\delta \mathcal{A}_{\text{Massless}}}{\delta W_{\alpha \dot{\alpha}}} = \overline{W}^{\alpha \dot{\alpha}} - \partial^{\alpha \dot{\alpha}} \overline{J}' - Z_R^{\dot{\alpha}} C^\alpha - \overline{Z}_L^\alpha \overline{C}^{\dot{\alpha}} \tag{57}$$

$$\delta \overline{J}' = \frac{\delta \mathcal{A}_{\text{Massless}}}{\delta \eta'} = \frac{1}{2} \left( \overline{\phi}_L^\delta C_\delta + \phi_R^{\dot{\delta}} \overline{C}_{\dot{\delta}} \right) - \overline{\Sigma}^{\beta \dot{\beta}} C_\beta \overline{C}_{\dot{\beta}} \tag{58}$$

We need to add the SUSY parts of $\delta$ to the above:
$$\delta_{\text{SUSY}} = \delta_\xi + \xi^{\alpha\dot\alpha}\partial_{\alpha\dot\alpha} \tag{59}$$
Adding these to the above ensures that $\delta^2 = 0$ for the entire operator.

**13**. In [16], it was shown[10] that the BRST cohomology of the operator $\delta_\xi$ in equation (1) was generated by the three basis vectors:
$$1, (C\xi)_{\dot\alpha}, (C\xi^2 C) \tag{60}$$
which can be multiplied by
$$C_{\alpha_1}\cdots C_{\alpha_n}. \tag{61}$$
There are also the complex conjugates
$$1, (\overline{C}\xi)_\alpha, (\overline{C}\xi^2\overline{C}) \tag{62}$$
which can be multiplied by the complex conjugates
$$\overline{C}_{\dot\alpha_1}\cdots \overline{C}_{\dot\alpha_n}. \tag{63}$$
Finally there is the real quantity
$$(C\xi\overline{C}) \tag{64}$$
which cannot be multiplied by any $C$ or $\overline{C}$ factors.

**14**. The Grading determines that:
$$\delta_{0\chi} = \int d^4x \left\{ \partial^{\alpha\dot\alpha}\overline{\omega}\frac{\delta}{\delta \overline{V}^{\alpha\dot\alpha}} - \frac{1}{g}\left[\partial_{\alpha\dot\alpha}V^{\alpha\dot\alpha} + 2J\right]\frac{\delta}{\delta\eta} \right. \tag{65}$$
$$+\partial_{\alpha\dot\alpha}\overline{\chi}_R^\alpha\frac{\delta}{\delta U_{L\dot\alpha}} + \partial_{\alpha\dot\alpha}\chi_L^{\dot\alpha}\frac{\delta}{\delta \overline{U}_{R\alpha}} - 2B\frac{\delta}{\delta\Xi} + \left(\partial^{\gamma\dot\delta}\Omega_{\gamma\dot\delta} - \Box\eta\right)\frac{\delta}{\delta K} + \Box\omega\frac{\delta}{\delta J} \tag{66}$$
$$\left. + \left(\Box\overline{V}_{\alpha\dot\alpha} + \frac{1}{2}\partial_{\alpha\dot\alpha}\partial^{\beta\dot\beta}\overline{V}_{\beta\dot\beta} + \partial_{\alpha\dot\alpha}\frac{1}{2g}\left(\partial_{\beta\dot\beta}\overline{V}^{\beta\dot\beta} + 2\overline{J}\right)\right)\frac{\delta}{\delta\Omega_{\alpha\dot\alpha}} \right\} + * \tag{67}$$

Using the Hodge decomposition technique of [15], we can deduce that the following (and their Complex Conjugates):
$$\eta, L, U_{L\dot\alpha}, U_{R\dot\alpha}, B, \Xi, K, \Omega_{\alpha\dot\alpha}, V_{\alpha\dot\alpha}, J, \omega_{\alpha\dot\alpha}, \Delta, V^{\alpha\dot\alpha}_{\alpha\dot\alpha}, \cdots \tag{68}$$
do not appear in $E_1$, but the following (and their Complex Conjugates) do appear in $E_1$ for the $\chi$ Sector:
$$\omega, \chi_{R\dot\alpha}, \chi_{L\dot\alpha}, G_{(\dot\alpha\dot\beta)} = V^\alpha_{\alpha(\dot\alpha\dot\beta)}, G_{(\alpha\beta)} = V^{\dot\alpha}_{\dot\alpha(\alpha\beta)}, \chi_{L\alpha(\dot\alpha\dot\beta)}, \chi_{R\alpha(\dot\alpha\dot\beta)} \cdots \tag{69}$$
Here $\chi_{R\alpha(\dot\alpha\dot\beta)} \equiv \partial_{\alpha(\dot\alpha}\chi_{R\dot\beta)}$, for example, is treated as a new kind of independent variable as explained in [15]. This one has dimension $\text{Dim}\, \chi_{R\alpha(\dot\alpha\dot\beta)} = \frac{5}{2}$.

---

[10]More general results, for other signatures and dimensions of space-time, can be found in [23,24,25,26] and [27], but here we use the form of the details in [16].

**15**. The Grading also determines that:

$$\delta_{0\,\phi} = \int d^4x \partial_{\gamma\dot{\delta}} W^{\gamma\dot{\delta}} \frac{\delta}{\delta\eta'} + \partial^{\alpha\dot{\alpha}} \overline{\phi}_{R\alpha} \frac{\delta}{\delta Z_L^{\dot{\alpha}}} \tag{70}$$

$$+ \partial^{\alpha\dot{\alpha}} \overline{\phi}_{L\alpha} \frac{\delta}{\delta Z_R^{\dot{\alpha}}} - \frac{1}{2} \Box E \frac{\delta}{\delta \Upsilon} + \left(-\partial^{\alpha\dot{\alpha}} J' + W^{\alpha\dot{\alpha}}\right) \frac{\delta \mathcal{A}}{\delta \Sigma_{\alpha\dot{\alpha}}} + * \tag{71}$$

Using the Hodge decomposition technique of [15], we can deduce that the following (and their Complex Conjugates):

$$\eta', Z_{L\dot{\alpha}}, Z_{R\dot{\alpha}}, \Upsilon, \Sigma_{\alpha\dot{\alpha}}, \partial_{\gamma\dot{\delta}} W^{\gamma\dot{\delta}}, \left(-\partial^{\alpha\dot{\alpha}} J' + W^{\alpha\dot{\alpha}}\right), \Box E, \cdots \tag{72}$$

do not appear in $E_1$ but the following (and their Complex Conjugates) can appear in $E_1$ for the $\phi$ Sector:

$$E, J', \phi_{R\dot{\alpha}}, \phi_{L\dot{\alpha}}, J'_{\alpha\dot{\alpha}} + W_{\alpha\dot{\alpha}}, E_{\alpha\dot{\alpha}}, \phi_{L\alpha(\dot{\alpha}\dot{\beta})}, \phi_{R\alpha(\dot{\alpha}\dot{\beta})}, E_{(\alpha\beta),(\dot{\alpha}\dot{\beta})}, \cdots \tag{73}$$

**16**. **Construction of $E_1$ with $m, m^2$:** Now we can construct the relevant subspace of $E_1$. It consists of all possible Lorentz invariant expressions with dimension Dim $=0$ and Lepton Number L$=0$, and one or two powers of mass, two Fields or Ghosts from the above sets in equations (69) and (73), and the cohomology results for $\delta_\xi$, as summarized above in equations (60) through (64) above. However, before we write $E_1$ down, we consider the effect of $d_1$ on $E_1$, and we will incorporate that into our results for $E_1$.

**17**. **Construction of $d_1$:** The operator $d_1$ maps various vectors in $E_1$ to other vectors in $E_1$. It can be written in the sectors that we look at here in the form

$$d_1 = \Pi_1 \left\{ G_{\alpha\beta} C^\beta \overline{\chi}^\dagger_{L\alpha} + \overline{G}_{\alpha\beta} C^\beta \overline{\chi}^\dagger_{R\alpha} + G_{\dot{\alpha}\dot{\beta}} \overline{C}^{\dot{\beta}} \chi^\dagger_{L\dot{\alpha}} + \overline{G}_{\dot{\alpha}\dot{\beta}} \overline{C}^{\dot{\beta}} \chi^\dagger_{R\dot{\alpha}} \right.$$

$$+ (\overline{C} \cdot \phi_R - C \cdot \overline{\phi}_L) \overline{E}^\dagger - \frac{1}{2}(C \cdot \overline{\phi}_L + \overline{C} \cdot \phi_R) \overline{J}'^\dagger$$

$$\left. + (C \cdot \overline{\phi}_R - \overline{C} \cdot \phi_L) E^\dagger - \frac{1}{2}(\overline{C} \cdot \phi_L + C \cdot \overline{\phi}_R) J'^\dagger \right\} \Pi_1 \tag{74}$$

where $\Pi_1$ is the orthogonal projection operator onto the space $E_1$. So for example, using the conclusion that the terms (72) are not in $E_1$, we see that

$$\Pi_1 \Sigma_{\alpha\dot{\alpha}} = 0 \tag{75}$$

and so this does not contribute to $d_1$.

**18**. To see what happens here it is useful to write the terms as functions of new variables defined as follows

$$E_+ = \frac{-1}{2}(E + 2J'); \ E_- = \frac{1}{2}(E - 2J'), \tag{76}$$

because these have simple maps under $d_1$. From (74) it follows that:

$$d_1 E_+ = \overline{C} \cdot \phi_L \Leftrightarrow d_1 \overline{E}_+ = C \cdot \overline{\phi}_L \tag{77}$$

$$d_1 E_- = C \cdot \overline{\phi}_R \Leftrightarrow d_1 \overline{E}_- = \overline{C} \cdot \phi_R \tag{78}$$

**19. Terms in $E_1$ with $m^2$:** Renormalize $N_G \to N_G - 4$ for the expressions, so that the action has Ghost Charge zero. We find that there is nothing in $E_1$ with Ghost Charge $G \geq 4$ or $G \leq 0$ with a factor $m^2$. Between those extremes there are the following $1 + 12 + 8 = 21$ objects, which divide into three sets. The maps made by $d_1$ do not connect the three sets. If $d_1$ maps an object in $E_1$ to another object in $E_1$, then both are killed, and neither object survives to $E_2$. This mutual murder is called the Elizabethan Drama. Here is a quote from J. F. Adams, author of a text on algebraic topology[11]

*...the behavior of this spectral sequence... is a bit like an Elizabethan drama, full of action, in which the business of each character is to kill at least one other character, so that at the end of the play one has the stage strewn with corpses and only one actor left alive (namely the one who has to speak the last few lines)...*

The remaining actors form the cohomology space.

**20. $\chi\chi$ Terms in $E_1$ with $m^2(C\xi\overline{C})$:** The following is not affected by $d_1$, and so it remains in $E_2$.
$$m^2(C\xi\overline{C})\omega\overline{\omega} \in E_2 \tag{79}$$
This object is real (if one multiplies by a factor of $i$).

**21. $\chi\chi$ Terms in $E_1$ with $m^2(C\xi^2 C)$:** Here there are 4 objects, but now they kill each other with mappings as follows:
$$m^2(C\xi^2 C) C \cdot \overline{\chi}_R \,\overline{\omega}, \xrightarrow{d_1} m^2(C\xi^2 C) C^\alpha G_{\alpha\beta} C^\beta \overline{\omega} \tag{80}$$
$$m^2(C\xi^2 C) C \cdot \overline{\chi}_L \,\omega \xrightarrow{d_1} m^2(C\xi^2 C) C^\alpha \overline{G}_{\alpha\beta} C^\beta \omega \tag{81}$$
Of course the same happens with their Complex Conjugates:
$$m^2(\overline{C}\xi^2\overline{C})\overline{C} \cdot \chi_L \,\overline{\omega} \xrightarrow{d_1} m^2(\overline{C}\xi^2\overline{C})\overline{C}^{\dot\alpha} G_{\dot\alpha\dot\beta} \overline{C}^{\dot\beta}\overline{\omega} \tag{82}$$
$$m^2(\overline{C}\xi^2\overline{C})\overline{C} \cdot \chi_R \,\omega \xrightarrow{d_1} m^2(\overline{C}\xi^2\overline{C})\overline{C}^{\dot\alpha} \overline{G}_{\dot\alpha\dot\beta} \overline{C}^{\dot\beta}\omega \tag{83}$$
As a result, none of these 8 objects survive to $E_2$.

**22. $\phi\chi$ Terms in $E_1$ with $m^2(C\xi^2 C)$:** Here we have six objects. We need to write these in terms of the variables (76), because these have simple maps under $d_1$ as shown in (77) and (78). The first two map to other objects, which eliminates four objects:
$$m^2(C\xi^2 C) E_- \overline{\omega} \xrightarrow{d_1} m^2(C\xi^2 C) C \cdot \overline{\phi}_R \,\overline{\omega} \tag{84}$$
$$m^2(C\xi^2 C) \overline{E}_+ \omega \xrightarrow{d_1} m^2(C\xi^2 C) C \cdot \overline{\phi}_L \,\omega \tag{85}$$
However the following two objects do not map to other objects, because of (77) and (78), and because $(C\xi^2 C)\overline{C}_{\dot\alpha}$ is not in the cohomology space of $\delta_\xi$, as summarized above in equations (60) through (64). So they remain in $E_2$:
$$m^2(C\xi^2 C) E_+ \overline{\omega} \in E_2 \tag{86}$$

---
[11]This quote comes (at second hand) from the useful text [28].

$$m^2(C\xi^2 C)\overline{E}_-\omega \in E_2 \tag{87}$$

Note that the mapping in this sector has dimensions:

$$4 \xrightarrow{d_1} 2 \tag{88}$$

and so it is clear that 2 of the objects do not get mapped, and therefore they remain in $E_2$. With the Complex Conjugates of these equations, there are 12 objects in all in this sector. From the 12 objects in this set, we find that four of them survive to $E_2$, and eight are killed.

23. **Summary of $E_2$ with $m^2$:** So of the $1+8+12 = 21$ objects with a factor of $m^2$, one real and two complex objects survive in equations (79) (86) and (87). In fact $E_2 = E_\infty$ here. This is obvious because all the $m^2$ objects in $E_2$ have ghost charge one, so all $d_r = 0, r \geq 2$ for this $m^2$ sector[12]. So these become equations (9), (11) and (12).

24. **Terms in $E_1$ with $m$:** There is nothing with Ghost Charge $G \geq 5$ or $G \leq -1$ with a factor $m$. We find that there are $4+18+24+8 = 54$ terms with the factor $m$, which divide into four sets. The maps made by $d_1$ do not connect the four sets. Again we examine the Elizabethan Drama.

25. **$\chi\phi$ Terms in $E_1$ with $m(C\xi\overline{C})$:** There are two terms here plus their complex conjugates, making a total of four terms. These do not map under $d_1$ and remain in $E_2$:

$$m(C\xi\overline{C})E\overline{\omega} \in E_2 \tag{89}$$

$$m(C\xi\overline{C})J'\overline{\omega} \in E_2, \tag{90}$$

and their complex conjugates are also in $E_2$:

$$m(C\xi\overline{C})\overline{E}\omega \in E_2,\ m(C\xi\overline{C})\overline{J}'\omega \in E_2 \tag{91}$$

26. **$\phi\phi$ Terms in $E_1$ with $m(C\xi^2 C)$:** There are 9 of these. This one survives because there are no $(C\xi^2 C)\overline{C}_{\dot\alpha}$ terms in the cohomology of $\delta_\xi$. :

$$m(C\xi^2 C)E_+\overline{E}_- \in E_2 \tag{92}$$

These two objects each map to one term:

$$m(C\xi^2 C)E_+\overline{E}_+ \xrightarrow{d_1} m(C\xi^2 C)\overline{E}_+(C\cdot\overline{\phi}_L) \tag{93}$$

$$m(C\xi^2 C)E_-\overline{E}_- \xrightarrow{d_1} m(C\xi^2 C)\overline{E}_-(C\cdot\overline{\phi}_R) \tag{94}$$

and so all four of these are killed.

The following object maps to a sum of two terms:

$$m(C\xi^2 C)E_-\overline{E}_+ \xrightarrow{d_1} m(C\xi^2 C)(C\cdot\overline{\phi}_R)\overline{E}_+ \tag{95}$$

$$+ m(C\xi^2 C)E_-(C\cdot\overline{\phi}_L) \tag{96}$$

---
[12] Recall that $\delta$ and $d_r$ both must increase the value of $N_G$ by 1, because of (25).

Then the other linear combination gets mapped as follows:

$$m(C\xi^2 C)(C \cdot \overline{\phi}_R)\overline{E}_+ - m(C\xi^2 C)E_-(C \cdot \overline{\phi}_L) \xrightarrow{d_1} \tag{97}$$

$$2m(C\xi^2 C)(C \cdot \overline{\phi}_R)(C \cdot \overline{\phi}_L) \tag{98}$$

Note that the mapping in this sector has dimensions:

$$1 \xrightarrow{d_1} 0 \tag{99}$$

$$3 \xrightarrow{d_1} 4 \xrightarrow{d_1} 1 \tag{100}$$

The 9 Complex Conjugates work the same way of course. So there are 18 terms of this kind in total, and two of them survive to $E_2$.

**27**. $\chi\phi$ **Terms in $E_1$ with $m(C\xi^2 C)$:** There are 12 terms here. This set kills itself off, and nothing survives in $E_2$. These two each map to one other object:

$$m(C\xi^2 C)(C \cdot \overline{\chi}_R)\overline{E}_- \xrightarrow{d_1} m(C\xi^2 C)(C \cdot G \cdot C)\overline{E}_- \tag{101}$$

$$m(C\xi^2 C)(C \cdot \overline{\chi}_L)E_+ \xrightarrow{d_1} m(C\xi^2 C)(C \cdot \overline{G} \cdot C)E_+ \tag{102}$$

The next two each map onto two objects:

$$m(C\xi^2 C)(C \cdot \overline{\chi}_R)\overline{E}_+ \xrightarrow{d_1} m(C\xi^2 C)(C \cdot \overline{\chi}_R)(C \cdot \overline{\phi}_L) \tag{103}$$

$$+ m(C\xi^2 C)(C \cdot G \cdot C)\overline{E}_+ \tag{104}$$

$$m(C\xi^2 C)(C \cdot \overline{\chi}_L)E_- \xrightarrow{d_1} m(C\xi^2 C)(C \cdot \overline{\chi}_L)(C \cdot \overline{\phi}_R) \tag{105}$$

$$+ m(C\xi^2 C)(C \cdot G \cdot C)E_- \tag{106}$$

Then the other linear combinations get mapped

$$m(C\xi^2 C)(C \cdot \overline{\chi}_R)(C \cdot \overline{\phi}_L) - m(C\xi^2 C)(C \cdot G \cdot C)\overline{E}_+ \tag{107}$$

$$\xrightarrow{d_1} 2m(C\xi^2 C)(C \cdot G \cdot C)C \cdot \overline{\phi}_L \tag{108}$$

$$m(C\xi^2 C)(C \cdot \overline{\chi}_R)(C \cdot \overline{\phi}_L) - m(C\xi^2 C)(C \cdot G \cdot C)\overline{E}_- \tag{109}$$

$$\xrightarrow{d_1} 2m(C\xi^2 C)(C \cdot \overline{G} \cdot C)C \cdot \overline{\phi}_R \tag{110}$$

Note that the mapping in this sector has dimensions:

$$4 \xrightarrow{d_1} 6 \xrightarrow{d_1} 2 \tag{111}$$

As a result nothing suvives in this sector of the $\chi\phi$ Terms in $E_1$ with $m(C\xi^2 C)$, or for its Complex Conjugate, which is another 12 terms.

**28**. $\chi\chi$ **Terms in $E_1$ with $m(C\xi^2 C)$:** There are 4 terms here. This set also kills itself off, and nothing survives in $E_2$.

$$m(C\xi^2 C)(C \cdot \overline{\chi}_L)(C \cdot \overline{\chi}_R) \xrightarrow{d_1} m(C\xi^2 C)(C \cdot G \cdot C)C \cdot \overline{\chi}_L \tag{112}$$

$$+ m(C\xi^2 C)(C \cdot \overline{G} \cdot C) C \cdot \overline{\chi}_R \tag{113}$$

and the other linear combination

$$m(C\xi^2 C)(C \cdot G \cdot C) C \cdot \overline{\chi}_L \tag{114}$$

$$- m(C\xi^2 C)(C \cdot \overline{G} \cdot C) C \cdot \overline{\chi}_R \tag{115}$$

$$\xrightarrow{d_1} m(C\xi^2 C)(C \cdot G \cdot C)(C \cdot \overline{G} \cdot C) \tag{116}$$

Note that the mapping in this sector has dimensions:

$$1 \xrightarrow{d_1} 2 \xrightarrow{d_1} 1 \tag{117}$$

So nothing survives out of this set of $\chi\chi$ Terms in $E_1$ with $m(C\xi^2 C)$, or its Complex Conjugate, which are another four terms.

**29. Summary of $E_2$ with $m$:** So out of all these $4 + 18 + 24 + 8 = 54$ '$m$ Terms', only 6 survive to $E_2$, namely (89) and (90) and (92) and their Complex Conjugates. In fact $E_2 = E_\infty$ again here. This is obvious because all the $m$ objects in $E_2$ have ghost charge zero, so all $d_r = 0, r \geq 2$ for these sectors[13]. These 6 terms yield (7), (8) and (10) in the summary at the beginning.

**30. Summary of map from $E_\infty \to \mathcal{H}$:** We will illustrate the map from $E_\infty \to \mathcal{H}$ with the item (9). Let us call it $h_0$:

$$h_0 = m^2 \omega \overline{\omega} (C\xi\overline{C}) \in E_\infty \Rightarrow \mathcal{O} \in \mathcal{H} \tag{118}$$

Using the full form of $\delta_0$ and $\delta_1$ above, we see that

$$\delta_0 h_0 = 0, \ \delta_1 h_0 = 0 \tag{119}$$

However the following is not zero:

$$\delta_2 h_0 = \left\{ -\overline{\omega}\xi \cdot \partial\omega - \omega\xi \cdot \partial\overline{\omega} + (CV\overline{C})\overline{\omega} - (C\overline{V}\,\overline{C})\omega \right\} m^2 (C\xi\overline{C}) \tag{120}$$

and if we define

$$h_2 = m^2 \left\{ (\xi V)\overline{\omega} + \omega(\xi \overline{V}) \right\} (C\xi\overline{C}) \tag{121}$$

then it is simple to verify that

$$\delta_2 h_0 + \delta_0 h_2 = 0 \tag{122}$$

We continue to build in this manner. The identity

$$\xi_{\alpha\dot\alpha}\xi_{\beta\dot\beta} = -\xi_{\beta\dot\beta}\xi_{\alpha\dot\alpha} = a_1 \varepsilon_{\alpha\beta}(\xi^2)_{(\dot\alpha\dot\beta)} + a_2 \varepsilon_{\dot\alpha\dot\beta}(\xi^2)_{(\alpha\beta)} \tag{123}$$

enables us to contract indices on multiple $\xi$ products. The numerical constants $a_i$ can be found by contracting (123). The spectral sequence guarantees that this process does not meet an obstruction. This construction can be done fairly carelessly, and one ends with an expression of the form

$$h = (\xi \cdots) + (\xi^2 \cdots) + (\xi^3 \cdots) + (\xi^4)\mathcal{I} \tag{124}$$

and then the last term gives us the corresponding object in $\mathcal{H}$:

$$h_0 \Rightarrow \mathcal{O} = \int d^4x \mathcal{I} \in \mathcal{H} \tag{125}$$

---
[13] Recall that $\delta$ and $d_r$ both must increase the value of $N_G$ by 1, because of (25).

Probably the best way to do this is to do enough of it to see the general form of $\mathcal{I}$, and then take the expression $\mathcal{O} = \int d^4x \mathcal{I}$ with free coefficients. This is how $\mathcal{A}_E$ in (5) was found. Using $\delta$ one can determine the coefficients so that $\mathcal{O} = \int d^4x \mathcal{I}$ is a cocycle of $\delta$ and not a coboundary. One can also check what terms are available to form a coboundary as we did above in equation (6), but it is not necessary to do so for most purposes.

**31**. From the above we can see that choosing $N_{\text{Grading}} = N_{\text{C}} + N_{\overline{\text{C}}} + 2N_\xi$ in (18), which implies that we use the cohomology of $\delta_\xi$, in $E_1$, causes the maps to become quite simple, and to close at $E_2$ in this example. The structure of SUSY is embedded in the calculation right at the beginning in $E_1$, which makes the beginning a bit of a chore, because much of the SUSY structure is incorporated at that stage. But then the rest of the calculation and the resulting form of $E_\infty$ are quite simple and fairly intuitive.

**32**. Most SUSY theories do not possess as much cohomology stucture as this SOSO model. The structure is clearly created to some extent by the presence of $J'$, which comes from the BRST recycling that created the massive SOSO superspin $\frac{1}{2}$ multiplet in [17]. The next challenge is to see whether the SOSO model can be coupled in a SUSY invariant way to other fields, so that it interacts, and that is a problem that can also be resolved using the above methods.

## Acknowledgments

I thank Carlo Becchi, Friedemann Brandt, Cliff Burgess, Philip Candelas,Rhys Davies, Mike Duff, Chris Hull, Pierre Ramond, Peter Scharbach, Kelly Stelle, Raymond Stora, Xerxes Tata, J.C. Taylor and Peter West for stimulating correspondence and conversations.